\documentclass[12pt]{iopart}
\pdfoutput=1

\usepackage{graphicx}
\usepackage{dcolumn}
\usepackage{bm}

\newcommand{\bq}{\begin{equation}}
\newcommand{\eq}{\end{equation}}
\newcommand{\bqa}{\begin{eqnarray}}
\newcommand{\eqa}{\end{eqnarray}}
\newcommand{\nn}{\nonumber \\}

\def\be     {\begin{equation}}
\def\ee     {\end{equation}}
\def\bea        {\begin{eqnarray}}
\def\eea        {\end{eqnarray}}
\def\bnn    {\begin{eqnarray*}}
\def\enn    {\end{eqnarray*}}

\def\k {\mathbf{k}}
\def\q {\mathbf{q}}
\def\Q {\mathbf{Q}}

\begin{document}

\title[Luttinger-Ward functional approach in the Eliashberg framework]{ Luttinger-Ward functional approach in the Eliashberg framework:\\
A systematic derivation of scaling for thermodynamics near the quantum critical point}
\author{A Benlagra$^{1,3}$, K-S Kim$^{2}$ and C P\'epin$^{1,4}$}
\address{1 Institut de Physique Th\'eorique, CEA, IPhT, CNRS,
URA 2306, F-91191 Gif-sur-Yvette, France} \address{2 Asia
Pacific Center for Theoretical Physics, Hogil Kim Memorial
building 5th floor, POSTECH, Hyoja-dong, Namgu, Pohang 790-784,
Korea}
\address{3 Institut f\"ur Theoretische Physik, Technische Universit\"at Dresden, 01062 Dresden, Germany.}
\address{4 International Institute of Physics, Universidade Federal do Rio Grande do Norte, 59078-400 Natal-RN, Brazil}
\date{\today}
\ead{adel.benlagra@mailbox.tu-dresden.de}

\begin{abstract}
Scaling expressions for the free energy are derived, using the Luttinger-Ward (LW) functional approach in the
Eliashberg framework, for two different models of quantum critical point
(QCP). First, we consider the spin-density-wave (SDW) model for which the 
effective theory is the Hertz-Moriya-Millis (HMM) theory, describing the interaction between  itinerant electrons and collective spin fluctuations. The dynamic of the latter are described by a dynamical exponent $z$ depending on the nature of the transition. Second,  we consider the Kondo breakdown model for QCP's, one
possible scenario for heavy-fermion quantum transitions, for which the effective theory is given by a gauge theory in terms of
conduction electrons, spinons for localized spins, holons for
hybridization fluctuations, and gauge bosons for collective spin
excitations. For both models, we construct the thermodynamic potential, in the whole phase diagram, including
all kinds of self-energy corrections in a self-consistent way, at the one loop level. We show how Eliashberg framework emerges at this level and use the resulting Eliashberg equations to simplify the LW expression for free energy . it is found that collective boson
excitations play a central role. The
scaling expression for the singular part of the free energy near
the Kondo breakdown QCP is characterized by two length scales : one is
the correlation length for hybridization fluctuations, and the
other is that for gauge fluctuations, analogous to the penetration
depth in superconductors.
\end{abstract}


\maketitle

\section{Introduction}
Fluctuation corrections are an essential ingredient near quantum critical points
(QCPs). It may be relatively easy to incorporate quantum
corrections in the weak coupling approach, the so called
Hertz-Moriya-Millis (HMM) theoretical framework \cite{HMM}.
However, it becomes more complicated to include quantum
fluctuations in the strong coupling approach such as the gauge
theoretical framework sometimes proposed to describe strongly
correlated electrons like doped Mott insulators
\cite{Gauge_Theory_Review} and some heavy-fermion QCPs
\cite{HF_GT}. For these models, it is believed that strong correlations fractionalize
electrons into some exotic elementary excitations carrying
fractional quantum numbers of electrons, and quantum fluctuations
of such enhanced degrees of freedom appear to be complicated. It
is challenging to develop a systematic approach to introduce, self-consistently, 
physically essential fluctuations into the thermodynamic potential near a QCP.
\newline

Effects of quantum corrections on the thermodynamic potential can
be incorporated systematically using the Luttinger-Ward (LW)
functional approach \cite{LW_Original,BK_Conservation, Coleman}, where the
grand potential is written in terms of dynamic quantities, such as the fully
dressed Green's function $G[\Sigma]$ and the self-energy $\Sigma$, through the relation\cite{Coleman}
\bqa \Omega[\Sigma] =  \mbox{T STr} \Bigl[ \ln \Bigl\{-
G^{-1}[\Sigma]\Bigr\} + \Sigma G[\Sigma] \Bigr] +
Y\Bigl\{G[\Sigma]\Bigr\} ,\label{LW} 
\eqa
where $\mbox{STr[A]=Tr[A$_B$]-Tr[A$_F$]}$ is the supertrace over Matsubara frequencies, internal quantum numbers of the bosonic (B) and fermionic (F) components of A. The quantity $Y\Bigl\{G[\Sigma]\Bigr\}$ is the so-called LW functional,
determined purely by the interaction potential and given by the sum of
all closed-loop two-particle irreducible skeleton diagrams in the
perturbation theory approach. Variation of the LW functional $Y$ with respect to $G$ generates the self-energy 
\bqa && \frac{\delta
Y\Bigl\{G[\Sigma_{s}]\Bigr\}}{\delta G[\Sigma_{s}]} =
\Sigma_{s} \equiv G_{0}^{-1} - G^{-1}[\Sigma_{s}]  \label{self}\eqa
where $G_0$ is the non-interacting Green's function.

The thermodynamic potential is stationary with respect to changes of the self-energy, i.e. it satisfies the saddle-point
condition
 \bqa && \frac{\delta \Omega
[\Sigma] }{\delta \Sigma } \Bigl|_{\Sigma = \Sigma_{s}} = 0 . \label{self_saddle}\eqa

An important issue in the perturbation approach of the LW
functional is to find an explicit functional dependence for $Y$\cite{Potthoff}. This is generally unknown and it is not always possible to sum the skeleton expansion into a closed form for $Y$. The problem with strongly correlated systems is even worse because the  convergence of the skeleton expansion is not guaranteed. It was demonstrated that the LW
functional can be written as a closed form in the Eliashberg
framework \cite{Chubukov_LW_FL,Chubukov_LW_FM}, where the
Eliashberg approximation allows to handle quantum corrections in a 
self-consistent way, at the one-loop level. The Eliashberg theory turns out
to be justified for $z > 1$ quantum criticality, where $z$ is the dynamical exponent,
 using a large-$N$ expansion supporting the Migdal theorem \cite{Chubukov_Migdal}. Here $N$
is the number of fermion flavors with spin symmetry $SU(2)$. In particular, the Eliashberg
theory was argued to be a minimal framework well working near a
QCP \cite{Chubukov_LW_FM}.
\newline

In this study, we derive a LW expression of the free energy for two models of itinerant QCPs : the spin-density-wave (SDW) model and the Kondo breakdown model.  This will allow us to describe thermodynamics near these QCPs starting from a microscopic model and incorporating self-consistently the effect of quantum fluctuations. It is the first time that this is done for the Kondo breakdown model.

The plan is as follows. In section \ref{SF}, we present the Spin-fermion (SF) model and derive in a systematic way the LW functional and show how Eliashberg equations for self-energies are derived. The expression of the free energy is simplified using Eliashberg equations and the scaling expression of its singular part is deduced. Section \ref{KB} is devoted to the Kondo Breakdown (KB) model. A particular care is taken to describe the Higgs part of the phase diagram. The effect of condensation is incorporated into a zero-order theory before considering a cumulant expansion in the fluctuations interaction. For this gauge theory, there are additional collective excitations, which results in the presence of two length scales in the scaling expression of the free energy. These two scales are related through the Anderson-Higgs mechanism. Section \ref{discussion} summarizes and discusses our main results. In particular,  theoretical structure differences between the HMM theory and the gauge theory of the Kondo breakdown QCP are emphasized.
Technical details are presented in the appendices.

\section{Review of the Luttinger-Ward functional approach in the
Eliashberg framework of the spin-fermion model}
\label{SF}
The standard model of quantum criticality in a metallic system is
the HMM theory. In this model, a dynamical exponent $z$, relating the
variation of the energy with the momentum $\omega \sim q^{z}$, characterizes the dynamics of collective excitations near the QCP.
In particular, $z = 3$ describes the ferromagnetic QCP while $z = 2$
describes the antiferromagnetic one. It is valuable to review
the construction of the LW functional in the HMM theoretical framework, discussed in the past\cite{Chubukov_LW_FL}, 
although several heavy fermion compounds have been shown not to
follow the $z = 2$ HMM theory
\cite{LGW_F_QPT_Nature,INS_Local_AF,dHvA,Hall,GR_Exp}. 
\subsection{Spin-fermion model}

We start from the so called spin-fermion model (SF)
\cite{Chubukov_LW_FM,Chubukov_QCP,Chubukov_dSC} for the SDW
transition \bqa&& \mathcal{S}_{SF} = T \sum_{k}
\psi_{\sigma k}^{\dagger} [- G_{0}^{-1}(k)] \psi_{\sigma k}  + \frac{1}{2} T \sum_{q} \chi_{0}^{-1}(q)
\vec{S}_q\cdot\vec{S}_{-q}  \nn &&+ g T^2\sum_{k}
\sum_{q} \psi_{\sigma\, k+q}^{\dagger}
\vec{\tau}_{\sigma\sigma'}\psi_{\sigma' k} \cdot \vec{S}_{-q} + \mathcal{O}[\{\vec{S}\}^{n};n \geq 3] , \label{SFM}\eqa
where we used the "relativistic" notation for energy-momentum $k\equiv (\k,i\omega)$, $q\equiv(\q,i\Omega)$ and the sum expression is defined as
\bqa && \sum_{k} ...
\equiv  \sum_{i\omega} \int_{|\k-\k_{F}|<\Lambda} \frac{d^{d}
k}{(2\pi)^{d}} ... , \,\,\,\,\,\,\,\, \sum_{q} ... \equiv 
\sum_{i\Omega} \int_{|\q-\Q|<\Lambda} \frac{d^{d} q}{(2\pi)^{d}} ...
\nonumber \eqa

In (\ref{SFM})$, \psi_{\sigma k}$ is the fermionic annihilation field for an electron with energy-momentum vector $k$ and spin $\sigma$, $\vec{S}_{q}$ is a bosonic field describing spin-fluctuations near a momentum $\Q$ and g is the coupling constant measuring the strength of the interaction between fermionic and bosonic excitations. The last term in (\ref{SFM}) stands for higher order terms in S. These are shown to be irrelevant for $d>2$ and marginal for $d=2$ and can therefore be neglected \cite{Chubukov_dSC}.

 In the absence of the interaction, fermionic and bosonic excitations are described by the bare electron Green's function $G_{0} (k)$ and
the bare spin susceptibility $\chi_{0}(q)$ respectively
 \bqa && G_{0}
(k) = \frac{z_{0}}{i\omega - v_{F} |\k - \k_{F}|} , \,\,\,\,\,\,\,\,  \chi_{0}(q) = \frac{\chi_{0}}{\xi_{0}^{-2} + |\q -
\Q|^{2} + \Omega^{2}/v_{s}^{2}} .\label{SF_bare_propagator}\eqa
  $z_{0}$ is the quasiparticle renormalization factor given by
the Fermi liquid theory, and the electron dispersion is linearized
with a Fermi velocity $v_{F}$ and Fermi momentum $\k_{F}$.  The electron's chemical potential can incorporate effects of the condensed part of the bosonic field. The
bare spin susceptibility is the usual Ornstein-Zernicke form where $\xi_0$ is the bare correlation length of spins and $\chi_{0}\xi^{2}_{0}$ is the static susceptibility. $v_{s}$ is the bare spin velocity.

The spin-fermion model Eq.(\ref{SFM}) is an effective low-energy model that can be
derived from the Hubbard-like model in the weak coupling
approximation \cite{Chubukov_LW_FM,Chubukov_QCP,Chubukov_dSC};
high-energy fermions, with energy above $\Lambda$, are integrated out to generate collective bosonic modes that mediate the interaction between fermions at energies smaller than $\Lambda$.  Dynamics of the low-energy fermions and the collective spin excitations are then described by Eq. (\ref{SF_bare_propagator}). 
\subsection{Eliashberg theory}
\label{eliashberg}
The Eliashberg framework allows the evaluation of the self-energies $\Sigma$ and $\Pi$, for electrons and spin fluctuations respectively, self-consistently assuming we can neglect the momentum dependence of the $\Sigma$ and vertex corrections. An extensive review of this technique is given in \cite{Chubukov_LW_FM}. We recall here the spirit of this technique and main results.

\begin{figure}[ht]
\centering
\includegraphics[width=3.5 in, angle = 0]{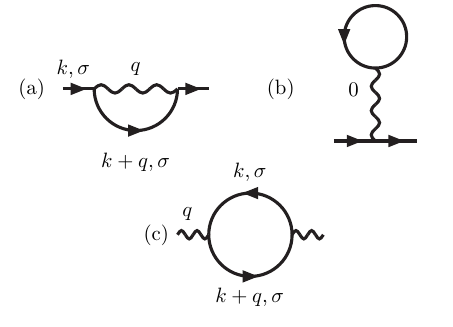}
\caption{(a) and (b) are the first order contribution to the fermionic self-energy and (c) is the polarization bubble, where $\sigma \in [1,N]$ is the spin index. The propagators of the fermion (straight line) and the spin fluctuations boson (wavy line) are fully dressed. (b) is a static and uniform part in the self-energy and can thus be considered as a renormalization of the electron chemical potential.}
\label{self_energies_SF}
\end{figure}

The Eliashberg procedure relies on three steps:

\begin{itemize}
\item neglect both the vertex corrections and the momentum dependence of the fermionic self-energy :
\bqa
\Sigma(\k, i\omega_n) = \Sigma(i\omega_n), \,\,\,\,\,\,\,\,
\Delta g = 0 \nonumber
\eqa
\item Use Dysons' equations 
\bqa && G^{-1}(\k, i\omega_n) = G_{0}^{-1}(\k, i\omega_n) -\Sigma(i\Omega_n)
 , \nn &&\chi^{-1}(\q, i\Omega_n)  = \chi_0^{-1}(\q, i\Omega_n) -\Pi(\q, i\Omega_n)
, \label{Dyson_SF}\eqa
to evaluate self-consistently the self-energies represented diagrammatically in Fig.- \ref{self_energies_SF}(a) and (c), where the propagators are fully dressed according to (\ref{Dyson_SF}). 
\item Check \textit{a posteriori} that the neglected momentum dependence of the fermionic self-energy and vertex corrections are indeed small. 
\end{itemize} 

The bosonic self-energy is found to be 
\bqa
\Pi(\q, i\Omega_n)&=&\gamma \frac{|\Omega_n|}{q^{z-2}},\label{damping_SF}
\eqa
where $\gamma = g^2 \chi_0 k_F/(\pi v_F^2)$ and $z$ is the dynamical exponent.  This result is customary for problems where fermions interact with their own collective modes. The latter are damped whenever they lie inside the particle-hole continuum of the Fermi liquid\cite{HMM}. Such a Landau-damped term is larger than the regular $\mathcal{O}(\Omega^2)$ term in the bare spin susceptibility Eq.(\ref{SF_bare_propagator}) and fully determines the collective spin dynamics. This causes feedback
effects on the self-energy correction of electrons, giving rise to
non-Fermi liquid physics near the QCP\cite{Chubukov_LW_FM}.
\newline

The model (\ref{SFM}) can be extended by introducing $N\neq 1$ identical fermionic species with spin symmetry $SU(2)$ \footnote{A problem in using the $SU(N)$ representation arises for the definition of the spin operator which is only possible in the $Sp(N)$ representation. This is a crucial aspect in particular when considering the spin fluctuations as critical modes. Then either we use the $Sp(N)$ representation or $N$ copies of $SU(2)$ fermions.}. A channel index $\nu \in  [1,N]$ is then added to the fermionic operators $\psi$ in (\ref{SFM}) and $g\rightarrow g/ \sqrt{N}$ to ensure a well-defined large N limit.

 It has been shown that the Eliashberg approximation becomes exact in the limit $N\rightarrow \infty$\cite{Chubukov_LW_FL, Chubukov_QCP}.  Indeed, it is shown that both vertex corrections and the momentum-dependent corrections to the fermionic self-energy turn out to scale as $1/N$ and vanish in the limit $N\rightarrow \infty$. 
 This limit shares some similarity with the Migdal limit for the electron-phonon problem : at large $N$, the damping introduced in (\ref{damping_SF}) scales as $N$ and the collective excitations become slow.  Then, the smallness in $1/N$ compares to the smallness in $m/M$ where $m$ is the electron's mass and $M$ is the ion's mass.
 
 In the following, we will show how the Eliashberg framework emerges from the LW approach to the Spin-Fermion model.

\subsection{Luttinger-Ward functional for the spin-fermion model}
\label{SF_LW}

\begin{figure}[ht]
\centering
\includegraphics[width=4.5 in, angle = 0]{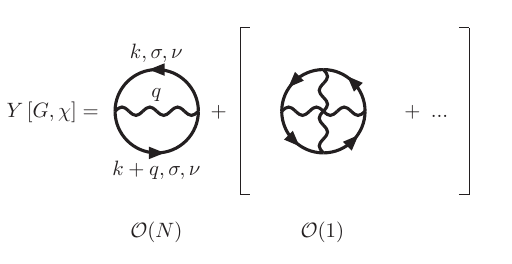}
\caption{Leading skeleton diagrams participating to the LW functional $Y$ for the Spin-Fermion model with dependence on $1/N$. Fermion (straight line) and boson (wavy line) propagators are fully dressed, $\sigma \in [1,N]$ is the spin index and $\nu \in [1,N]$ is the channel index. The first diagram contains one fermionic loop carrying spin and channel quantum numbers and one pair of vertices, each of order $\mathcal{O}(1/\sqrt{N})$, so that it is of order $\mathcal{O}(N^2/N)=\mathcal{O}(N)$. The second diagram involves one fermionic loop and two pairs of vertices so that it is of order $\mathcal{O}(N^2/N^2)=\mathcal{O}(1)$. Bracketed terms are dropped in the large $N$ limit.}
\label{Y_SF}
\end{figure}

As said in the introduction, the LW functional is, diagrammatically,  the sum of all closed-loop two-particle irreducible skeleton diagrams\cite{LW_Original}. These can be ordered in a $1/N$ expansion as in Fig.-\ref{Y_SF}

Considering the general expression of the LW expression [Eq.(\ref{LW})], and taking into account only the leading $\mathcal{O}(N)$ contribution to $Y$, shown in Fig.-\ref{Y_SF} and derived in \ref{app:derivation_LW}, 
one can write down the free energy in terms of electron and
spin-fluctuation Green's functions and self-energies 
\bqa 
\hspace{-1.5cm}F_{LW}[\Sigma,\Pi] &=& -  N T
\sum_{k} 
\Bigl[ \ln\Bigl\{ - G^{-1}(k)\Bigr\} +
\Sigma(k) G(k) \Bigr] + T \sum_{q} \Bigl[ \ln\Bigl\{
\chi^{-1}(q)\Bigr\}  +\Pi(q)
\chi(q) \Bigr] \nn & +& 3 N g^{2} T^2 \sum_{k,q}
G(k) \chi(q)G(k+q) , \label{LW_SF}\eqa where $G(k)$ and
$\Sigma(k)$ are the fully renormalized electron Green's function
and self-energy, while $\chi(q)$ and $\Pi(q)$ are the 
fully renormalized spin-fluctuation Green's function and
self-energy. The last term in (\ref{LW_SF}) corresponds to the leading skeleton diagram of order $\mathcal{O}(N)$ shown in Fig.-\ref{Y_SF}.

\subsection{Eliashberg equations}
\label{ss:eliashberg_SF}
One of the important aspects of the LW functional approach is that we can recover the self-consistent Eliashberg equations for self-energies. Indeed, if we restrict ourselves to the leading $\mathcal{O}(N)$ term in $Y$ and use the stationarity of the free energy (\ref{LW_SF}) with respect to self-energies (\ref{self_saddle}), we get the following equations
 \bqa && 
\frac{\delta G}{\delta \Sigma}\left ( - \Sigma(k) + 3 g^2 T \sum_{q}G(k+q)\chi(q) \right ) = 0 , \nn && 
\frac{\delta \chi}{\delta \Pi} \left ( \Pi(q) + 3 N g^2 T \sum_{k}G(k+q)G(k)\right )
= 0 , \nonumber \eqa 
from which we deduce immediately the expressions of the electronic self-energy and the collective spin polarization
 \bqa  \Sigma(k)& = &3g^{2} T \sum_{q} G(k+q) \chi(q) , \nn
 \Pi(q)& = &- 3 N g^{2} T
\sum_{k} G(k+q) G(k). \label{selfs_SF} \eqa 

\begin{figure}[ht]
\center
\includegraphics[width=4 in, angle = 0]{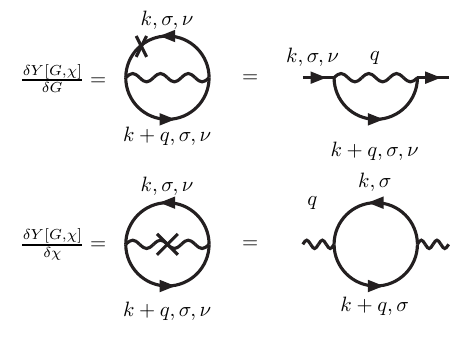}
\caption{Illustrating functional derivative of the LW functional with respect to Green's functions. The cross indicates the line that is cut by functional differentiation.}
\label{functional_sf}
\end{figure}

These expressions can be also obtained by differentiating the leading order $\mathcal{O}(N)$ contribution to the LW functional $Y$ with respect to $G$ and $\chi$ respectively, according to equation (\ref{self}). Diagrammatically, this is equivalent to cutting one of the internal lines of the corresponding diagram, as shown in Fig.-\ref{functional_sf}.
\newline

Equations (\ref{selfs_SF}), with Dyson's equations, are nothing but the self-consistent Eliashberg equations for self-energies (see Fig.\ref{self_energies_SF}).

\begin{figure}[ht]
\center
\includegraphics[width=1.7 in, angle = 0]{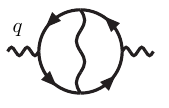}
\caption{Spin-fluctuations self-energy generated from the contribution to the LW functional of order $\mathcal{O}(1)$, shown in Fig.\ref{Y_SF}, by cutting one internal bosonic line. This can be obtained from The Eliashberg form of the spin-fluctuations polarization shown in Fig.-\ref{self_energies_SF}(c)  by inserting a bosonic propagator to the vertex.}
\label{vertex_correction_sf}
\end{figure}

Considering further terms in the LW functional $Y$ amounts to studying deviations from the Eliashberg theory, in particular introducing vertex corrections as shown in Fig.-\ref{vertex_correction_sf}, in the $1/N$ expansion. 

\subsection{Simplification of the Luttinger-Ward expression}
One can simplify further the full expression  Eq.(\ref{LW_SF}) of the free energy \cite{Chubukov_LW_FL}. Indeed, from  equations (\ref{selfs_SF}), one can notice that

\bqa
Y  \left \{ G, \chi \right \} & \equiv & 3 N g^{2} T^2 \sum_{k,q}G(k) \chi(q)G(k+q) \nn 
&=& N T \sum_{k} \Sigma(k)G(k) \label{Y_FSE_SF}\\
& = & - T \sum_{q} \Pi(q)\chi(q)\label{Y_BSE_SF}.
\eqa
Thus, if we insert Eq. (\ref{Y_BSE_SF}) into Eq.(\ref{LW_SF}), the latter reduces to  

\bqa  F_{eff}& =& - N T
\sum_{k} 
\Bigl[ \ln\Bigl\{ - G^{-1}(k)\Bigr\}  +
\Sigma(k) G(k) \Bigr] + T \sum_{q}
\ln\Bigl\{\chi^{-1}(q)\Bigr\} , \label{free_energy_F_SF}
\eqa
while if we insert Eq. (\ref{Y_FSE_SF}) we get the following expression for Eq.(\ref{LW_SF})

\bqa  F_{eff}& =& - N T
\sum_{k} \ln\Bigl\{ - G^{-1}(k)\Bigr\}  +T \sum_{q}\Bigl[
\ln\Bigl\{\chi^{-1}(q)\Bigr\}+\Pi(q) \chi(q)\Bigr] .   \label{free_energy_B_SF}\eqa

Considering equation (\ref{free_energy_F_SF}), we can show (See \ref{app:fermi_HMM}) that the fermionic part reduces to a Fermi liquid form $F_{FL} \equiv - \frac{\pi N\rho_{F} }{6} T^{2}$ so that the final expression of the free
energy for thermodynamics in the Eliashberg framework writes \bqa &&
F_{eff} = F_{FL} + T \sum_{q} \ln\Bigl\{ \chi^{-1}(q) \Bigr\}. \label{free_energy_SF} \eqa 

\subsection{Thermodynamics}
Performing the energy and momentum integrals in the Eliashberg
equations Eq. (\ref{selfs_SF}), one finds \cite{Chubukov_QCP,Chubukov_dSC}
\bqa && \Pi(\q,i\Omega) = \gamma \frac{|\Omega|}{q^{z-2}} , \eqa where
$\gamma = N g^2 \chi_0 k_F/(\pi v_F^2)$.
Considering only the linear-frequency Landau term in the spin susceptibility, this writes\bqa
&& \chi^{-1}(\q,i\Omega) =  \chi_{0}^{-1} \Bigl( \xi^{-2} +
|\q-\Q|^{2} + \frac{|\Omega|}{q^{z-2}} \Bigr) . \eqa 
In particular, we have $z=2$ for an antiferromagnetic QCP and $z=3$ for a ferromagnetic one.

Using this expression, the singular part of the free energy (\ref{free_energy_SF}) writes \bqa &&
f_{s}(\xi^{-2},T) = T \sum_{i\Omega, q} \ \ln\Bigl\{ \xi^{-2} +
|\tilde{q}|^{2} + \gamma \frac{|\Omega|}{q^{z-2}} \Bigr\} \nn && = -
\frac{1}{\pi} \int_{0}^{\infty} d \nu
\coth\Bigl(\frac{\nu}{2T}\Bigr) \int \frac{d^d \tilde{q}}{(2\pi)^d} \tan^{-1} \Bigl( \frac{\gamma\nu/\tilde q^{z-2}}{\xi^{-2} +
\tilde{q}^{2}} \Bigr) , \nn \label{singular_LW_SF}\eqa where $\tilde{\q} = \q - \Q$ is the
shifted momentum near the wave vector $\Q$.
Performing the frequency and momentum integrals in this equation,
one finds the analytic expression of (\ref{singular_LW_SF}). Details of this evaluation for $d=3, z=2$ are given in appendix \ref{app:singular_SF}.

We would like to emphasize that the resulting effective free energy
satisfies the following scaling relation \bqa && f_{s}(r,T)
= b^{-(d+z)} f_{r}(r b^{1/\nu}, T b^{z}) , \label{scaling_SF_1}\eqa where
$f_{r}(x,y)$ is an analytic regular function. $r \sim \xi^{-2}$ measures the distance to the QCP\footnote{Within the Eliashberg approximation, the correlation length's critical exponent value $\nu=1/2$, coming from the Ornstein-Zernicke form for the static boson propagator in Eqs. (5, 16) and valid at high energy, has no quantum corrections. This coincides with its mean-field value in the Hertz theory above the upper critical dimension, i.e. when $d+z>4$. }, $\nu$ is the correlation-length exponent, 
$z$ is the dynamical exponent and $d$ is the space dimension.
Inserting $b = \xi^{2\nu}$ into the above scaling expression, we
find \bqa && f_{s}(\xi^{-2},T) = \xi^{-2\nu(d+z)} f_{r}(1, T \xi^{2\nu
z}) . \label{scaling_SF_2}\eqa  Now, one can understand thermodynamics near
the HMM theory QCP based on this scaling free energy, derived from the
effective field theory in the Eliashberg framework.
\newline

As is already known, the naive scaling (\ref{scaling_SF_1}) may be spoiled by the presence of at least one dangerously irrelevant variable \cite{HMM} leading to a generalized scaling form
\be f(\xi^{-2}, T, u) = b^{-(d+z)}f(\xi^{-2} b^{1/\nu}, Tb^z, ub^{d+z-4}),\label{generalized_scaling}\ee
where $u$ is the constant coefficient of an additional $\phi^4$ term in the theory.

 In practice, for thermodynamics quantities, the results obtained from the naive scaling hold up to logarithmic corrections where the argument of the logarithm is a power of $T$. We comment in \ref{app:corrections_scaling} on how the effect of such variables can be handled within our method and show that we can obtain the same generalized expression (\ref{generalized_scaling}). The purpose in this section is only to show how one can get an analytic expression for free energy including quantum corrections in a self-consistent way.

\section{Luttinger-Ward functional in the Eliashberg framework of the Kondo breakdown
scenario}
\label{KB}

The HMM theoretical framework has been regarded as the
standard model for quantum criticality in metals for a long time, although several heavy
fermion compounds have been shown not to follow its predictions
\cite{LGW_F_QPT_Nature,INS_Local_AF,dHvA,Hall,GR_Exp}. An interesting alternative theory suggests that heavy fermion quantum transitions are selective Mott transitions of the $f$ nearly localized fermions \cite{Senthil_Vojta_Sachdev,Paul_KBQCP,Pepin_KBQCP,DMFT} at which the Kondo effect breaks down.  This scenario is supported by the presence of localized magnetic moments at the transition towards
magnetism \cite{INS_Local_AF}  and  Fermi surface
reconstruction at the QCP \cite{dHvA,Hall}.

This problem has been tackled using  the U(1) slave-boson
representation of the Anderson lattice model\cite{Paul_KBQCP,Pepin_KBQCP}, with the introduction of a small dispersion for the $f$-electrons. A remarkable aspect of the theory is that the resulting QCP, at which an effective hybridization vanishes, is multi-scale. Indeed, because we have two kind of fermions in the model, i.e the conduction $c$-fermions and the $f$-spinons, there exist a Fermi
surface mismatch $q^{*} = |k_{F}^{f} - k_{F}^{c}|$ between Fermi
momentum $k_{F}^{f}$ for spinons and $k_{F}^{c}$ for conduction
electrons since fillings of spinons and electrons differ from each
other. This mismatch gives rise to an energy gap $E^{*}$
for spinon-electron fluctuations that controls the dynamics of hybridization fluctuations. Although it depends on the
value of $q^{*}$, this energy scale is shown to vary from ${\cal
O}(10^{0})$ $mK$ to ${\cal O}(10^{2})$ $mK$. When $E < E^{*}$,
holon fluctuations are undamped, thus described by $z = 2$ dynamical exponent. On
the other hand, when $E > E^{*}$, holon fluctuations are
dissipative since spinon-electron excitations are Landau damped,
thus described by $z = 3$ critical theory. Based on the $z = 3$
quantum criticality, recent studies have found quasi-linear
electrical transport and logarithmically divergent specific heat
coefficient in $d = 3$ \cite{Pepin_KBQCP,Paul_KBQCP,Kim_Pepin},
and a divergent Gr\"uneisen ratio with an anomalous exponent
$0.7$ \cite{Kim_Adel_Pepin}, consistent with experiments
\cite{LGW_F_QPT_Nature,GR_Exp}. 

\subsection{U(1) slave-boson representation of the Anderson lattice model}

We start from the Anderson lattice model in the large-$U$ limit
\bqa && \mathcal{L} = \sum_{i}c_{i\sigma}^{\dagger}\left ( (\partial_{\tau}-\mu)\delta_{ij} - t_{ij} \right )c_{j\sigma} +
\sum_{i}d_{i\sigma}^{\dagger}(\partial_{\tau} +
\epsilon_{f})d_{i\sigma} \nn && + V \sum_{i}
(d_{i\sigma}^{\dagger}c_{i\sigma} + H.c.)   + J \sum_{\langle ij \rangle}
\vec{S}_{i}\cdot\vec{S}_{j} ,\label{large-U} \eqa where $c_{i\sigma}$ and
$d_{i\sigma}$ are conduction electron with a chemical potential
$\mu$ and localized electron with an energy level $\epsilon_{f}$,
respectively, $t_{ij}$ the hopping term of the conduction electron and $V$ the hybridization between c- and d-electrons. The last spin-exchange term is generated by a perturbative expansion to second order in $t/U$ and is in competition with the hybridization term. 
\newline

In the $U \rightarrow \infty$ limit of (\ref{large-U}), the strong correlations between the d-electrons show as a constraint of no double occupancy for the d-electron. This can be handled using the U(1) slave-boson representation \bqa &&
d_{i\sigma} = b_{i}^{\dagger} f_{i\sigma}, \eqa
where $b_{i}$ and $f_{i\sigma}$ are holon and spinon,
associated with hybridization and spin fluctuations, respectively, obeying the local constraint \bqa && b_{i}^{\dagger}b_{i} +
\sum_{\sigma}f_{i\sigma}^{\dagger} f_{i\sigma} = S N , \label{constraint} \eqa where $S = 1/2$ is the value of spin  and $N$ is the
number of fermion flavors with $\sigma = 1, ..., N$. 

One can then
rewrite Eq. (\ref{large-U}) into
 \bqa
 &\mathcal{L} = \sum_{\langle i j \rangle}
c_{i\sigma}^{\dagger}\left ( (\partial_{\tau}-\mu)\delta_{ij} - t_{ij} \right )c_{j\sigma}  + \sum_{i} f_{i\sigma}^{\dagger}(\partial_{\tau} +
\epsilon_{f})f_{i\sigma}  + b_{i}^{\dagger} \partial_{\tau} b_{i} \nn &+ V \sum_{i} (b_{i}f_{i\sigma}^{\dagger}c_{i\sigma} +
H.c.)  + J \sum_{\langle ij \rangle} (
f_{i\sigma}^{\dagger}\chi_{ij}f_{j\sigma} + H.c.) \nn &+
 N J
\sum_{\langle ij \rangle} |\chi_{ij}|^{2} + i \sum_{i} \lambda_{i}
(b_{i}^{\dagger}b_{i} + f_{i\sigma}^{\dagger} f_{i\sigma} - S N)   \label{lagrangian_KB_b}
\eqa 
The spin-exchange term for the localized orbital has been decomposed, using a field $\chi_{ij}$,  resulting in
exchange hopping processes for the spinons. The local constraint (\ref{constraint}) is taken into account by the introduction of a Lagrange multiplier $\lambda_{i}$.
\newline

Performing the saddle-point approximation of $b_{i} \rightarrow
b$, $\chi_{ij} \rightarrow \chi$, and $i\lambda_{i} \rightarrow
\lambda$, one finds an orbital selective Mott transition as
breakdown of Kondo effect at $J \approx T_{K}$, where a
spin-liquid Mott insulator ($ b = 0$) arises in
$J > T_{K}$ while a heavy-fermion Fermi liquid ($
b \not= 0$) results in $T_{K} > J$
\cite{Senthil_Vojta_Sachdev,Paul_KBQCP,Pepin_KBQCP}. Here, $T_{K}
= D \exp\Bigl(\frac{\epsilon_{f}}{N \rho_{c}V^{2}}\Bigr)$ is the
Kondo temperature, where $\rho_{c} \approx (2D)^{-1}$ is the
density of states for conduction electrons with the half bandwidth
$D$.
\newline

\begin{figure}[ht]
\center
\includegraphics[width=1.7 in, angle = 0]{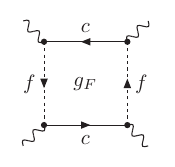}
\caption{Four-point electron-spinon polarization for the holons.}
\label{g4}
\end{figure}

Beyond the mean-field approximation, gauge fluctuations
corresponding to phase fluctuations of the hopping parameter
$\chi_{ij} = \chi e^{ia_{ij}}$ should be introduced to express
collective spin fluctuations. It is more convenient to represent
the above effective Lagrangian as follows, performing the
continuum approximation \footnote{The fermionic and bosonic fields here are all time and position dependent. A full discussion of this Lagrangian can be found in \cite{Pepin_KBQCP}.}, 
\bqa && {\cal L}_{ALM} =\sum_{\sigma} \int d\mathbf{r} \,\,\,\,
c_{\sigma}^{*}(\partial_{\tau} - \mu_{c})c_{\sigma} +
\frac{1}{2m_{c}}|\partial_{i} c_{\sigma}|^{2} +
f_{\sigma}^{*}(\partial_{\tau} - \mu_{f} -
ia_0)f_{\sigma} \nn &&+ \frac{1}{2m_{f}}|(\partial_{i} -
ia_{i})f_{\sigma}|^{2}  +b^{*}(\partial_{\tau} -
\mu_b -ia_0)b + \frac{1}{2m_{b}}|(\partial_{i} - ia_{i})b|^{2} + \frac{u_{b}}{2} |b|^{4} \nn &&+  V
(b^{*}c_{\sigma}^{*}f_{\sigma} + H.c.) +
\frac{1}{4g^{2}} F_{\mu\nu}F_{\mu\nu} + S N (\mu_{b} + i a_0)
, \eqa
where $a_0$ is the scalar gauge field, $a_i$ is the $i$ component of the vectorial gauge field $\vec a$, $g$ is an effective coupling
constant between matter and gauge fields, $F_{\mu\nu}\equiv \partial_\mu a_\nu-\partial_\nu a_\mu$ is the fictitious electromagnetic tensor associated with the four-potential $(a_0, \vec a)$. Furthermore, several quantities,
such as fermion band masses and chemical potentials, are redefined
as follows
 \bqa
 & \lambda \rightarrow - \mu_{b} , ~~~ (2m_{c})^{-1} = t , ~~~ (2m_{f})^{-1} = J \chi, \nn
 & \mu_{c} = \mu + 2d t , ~~~ - \mu_{f} = \epsilon_{f} +\lambda - 2 J d \chi.
\eqa
In here, fermion bare bands $\epsilon_{k}^{c}$ and
$\epsilon_{k}^{f}$ for conduction electrons and spinons,
respectively, are treated in the continuum approximation as
follows \bqa && \epsilon_{k}^{c} = - 2 t (\cos k_{x} + \cos k_{y}
+ \cos k_{z}) \approx - 2 d t + t (k_{x}^{2} + k_{y}^{2} +
k_{z}^{2}) , \nn && \epsilon_{k}^{f} = - 2 J \chi (\cos k_{x} +
\cos k_{y} + \cos k_{z})  \approx - 2 J d \chi + J \chi
(k_{x}^{2} + k_{y}^{2} + k_{z}^{2}).
\eqa
The band dispersion for hybridization can arise from
high energy fluctuations of conduction electrons and spinons.
Actually, the band mass of holons is given by $m_{b}^{-1} \approx
N V^{2} \rho_{c} /2$, where $\rho_{c}$ is the density of states
for conduction electrons \cite{Paul_KBQCP,Pepin_KBQCP}. Local
self-interactions denoted by $u_{b}$ can be introduced via
non-universal short-distance-scale physics. One physical process
for such interactions is four-point electron-spinon polarization (see Fig.- \ref{g4}),
giving rise to $u_{b} = u_{0} \frac{V^{4}}{D^{3}}$ with $u_{0}
\approx \mathcal{O}(1)$. Because such a local interaction term
results from non-universal physics, one may consider that this
term is introduced phenomenologically.

Maxwell dynamics for gauge fluctuations appears from high energy
fluctuations of spinons and holons. 
\newline

We would like to develop the LW functional approach, including the
Higgs or heavy-fermion phase. In this respect we write the holon
field with its condensation part and fluctuation contribution
separately, 
\bqa && b \rightarrow \mathcal{B} + b . \eqa 
Then, the effective continuum Lagrangian is written as follows 
\bqa
&& {\cal L}_{ALM} = \sum_{\sigma} \int d\mathbf{r} \,\,c_{\sigma}^{*}(\partial_{\tau}  + \frac{1}{2m_{c}}\partial_{i}^2-\mu_{c})c_{\sigma}   +f_{\sigma}^{*}(\partial_{\tau} - \mu_{f} -ia_0)f_{\sigma} \nn 
&&+ \frac{1}{2m_{f}}|(\partial_{i} -
ia_{i})f_{\sigma}|^{2}   + b^{*}[\partial_{\tau} - (\mu_{b}- 2 u_{b} \mathcal{B}^{2}) - ia_0]b +
\frac{1}{2m_{b}}|(\partial_{i} - ia_{i})b|^{2} \nn
&& +\frac{u_{b}}{2} |b|^{4} + V(b^{*}c_{\sigma}^{*}f_{\sigma} + H.c.) + V \mathcal{B}(c_{\sigma}^{*}f_{\sigma} + H.c.) \nn
&& + \frac{1}{4g^{2}} f_{\mu\nu}f_{\mu\nu} + \frac{\mathcal{B}^{2}}{2m_{b}} a_{i}^{2} +\frac{u_{b}}{2} \mathcal{B}^{4} + \Bigl( S N - \mathcal{B}^{2} \Bigr) (\mu_{b} + ia_0). \nn \label{field-lagrange}
\eqa

We see that the chemical potential for holon excitations is modified from $\mu_{b}$ to $\mu_{b} - 2 u_{b}\mathcal{B}^{2}$. An important point is that gauge fluctuations
become gapped when $\mathcal{B} \not= 0$, which is due to the
Anderson-Higgs mechanism.

\subsection{Luttinger-Ward functional in the Kondo breakdown
scenario}
In the following, we demonstrate how thermodynamics can be extracted from the
complicated effective field theory described by (\ref{field-lagrange}), where two kinds of fermion
excitations and two kinds of boson fluctuations are coupled with
each other. The point is how to introduce all self-energy
corrections self-consistently. As discussed before, we construct
the LW functional in the Eliashberg framework, allowing us to take
all kinds of self-energy corrections self-consistently at least in
the one-loop level. 

For simplicity, we start by ignoring gauge fluctuations corrections, considering only holon fluctuations. Gauge fluctuations are after that manipulated in the same way once their coupling with holons and spinons is known.

\subsubsection{Constructing a zero-order theory}

A subtle issue in deriving a LW expression for free energy is how to handle a non-vanishing condensation $\mathcal{B} \neq 0$ to describe the Higgs phase. A first step towards this derivation is to construct a "zero-order" theory taking into account, in a proper way, the effect of the condensation part $\mathcal{B}$.

Going to Fourier space, we can cast the action corresponding to the Lagrangian (\ref{field-lagrange}) into a mean-field part and holon fluctuations part \footnote{The contribution of the local interaction with strength $u_b$ to the LW functional is sub-leading with respect to the diagram of order $\mathcal O(N)$ used to derive the Eliashberg equations below. The corresponding term is thus dropped from $\mathcal S_{fluc}$. Its inclusion may result in a generalized scaling form for the free energy as shown in Appendix E.} :

\bqa
&&\mathcal{S}_{MF} = - T \sum_{k} \left [ c^{\dagger}_{\sigma k} g^{-1}_c(k) c_{\sigma k}+f^{\dagger}_{\sigma k}g^{-1}_f(k, i \omega) f_{\sigma k}\right ] \nn
&& + V\mathcal{B}T\sum_{k} \left ( f^\dagger_{\sigma k} c_{\sigma k} + H.c. \right ) +  \Bigl( S N - \mathcal{B}^{2} \Bigr) \mu_{b} + u_b\frac{\mathcal{B}^4}{2}\nn
&&\mathcal{S}_{fluc}= -T\sum_{q\neq 0}b^{\dagger}_{ q} d^{-1}_b(q) b_{q} + V T^2\sum_{k ,\sigma} \sum_{q \neq 0}\left (b_k f^\dagger_{\sigma k+q} c_{\sigma k} + H.c. \right ) ,
\eqa
where 
\bqa 
 &&g_{c}^{-1}(k) = i\omega + \mu_{c} - \frac{\k^{2}}{2m_{c}}  , \nn  &&
g_{f}^{-1}(k) = i\omega + \mu_{f} -
\frac{\k^{2}}{2m_{f}} , \nn  
&&d_b^{-1}(q)= i\Omega + \mu_{b} - 2 u_b \mathcal{B}^2- \frac{\q^{2}}{2m_{b}}  , \eqa 

The interaction term gives raise to two kind of vertices, shown in the following figures

\begin{figure}[ht]
\center
\includegraphics[width=3.5 in, angle = 0]{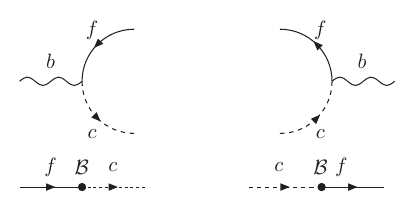}
\caption{ Vertices due to the interaction between fermions and holons. Here, a line stands for the spinon propagator, the dashed line for the electron propagator and the wavy line for the holon propagator. }
\label{vertices_b}
\end{figure}
Whereas it is justified to follow the same strategy as for the HMM framework (See Sec.\ref{SF_LW}), i.e. use the cumulant expansion to the second order, for the fluctuations part of the interaction term, it is not the case for the condensation part. However, the latter can be considered as a renormalization of the propagators $g_c$ and $g_f$ and is thus included in a new zero-order theory whose bare action is $\mathcal{S}_{MF}$.

 Indeed, we can write
\bqa\mathcal{S}_{MF} = - T \sum_{\mathbf{k}} ( c^{\dagger}_{\sigma \mathbf{k}} \,\, f^{\dagger}_{\sigma \mathbf{k}}) G_0^{-1}  \left ( \begin{array}{c}c_{\sigma \mathbf{k}}\\ f_{\sigma \mathbf{k}}\end{array}  \right )
+  \Bigl( S N - \mathcal{B}^{2} \Bigr) \mu_{b} + u_b \frac{\mathcal{B}^4}{2},
\eqa
where
$$G_0^{-1}=\left ( \begin{array}{cc}
g_{c}^{-1}& -V \mathcal{B}\\
-V \mathcal{B}& g_{f}^{-1}
\end{array} \right ).$$
This gives the renormalized matrix Green's function for the fermions
$$G_0=\left ( \begin{array}{cc}
G_{cc}^0& G_{cf}^0\\
G_{fc}^0& G_{ff}^0
\end{array} \right ),$$
where
\bqa
&&G_{ff}^0=\frac{g_c^{-1}}{g_c^{-1}g_f^{-1}-(V\mathcal{B})^2}, ~~~ 
G_{cc}^0=\frac{g_f^{-1}}{g_c^{-1}g_f^{-1}-(V\mathcal{B})^2} \nn
&&G_{fc}^0=G^0_{cf}=\frac{V \mathcal{B}}{g_c^{-1}g_f^{-1}-(V\mathcal{B})^2}
\eqa

\begin{figure}[ht]
\center
\includegraphics[width=4.1 in, angle = 0]{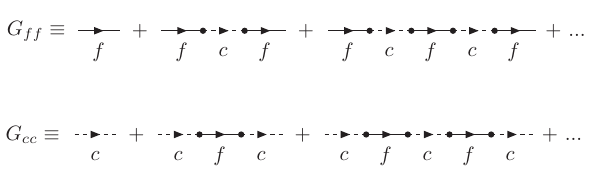}
\caption{ Propagators of the zero-order theory, where the effect of the condensation $\mathcal{B}$ is totally taken into account. }
\label{propagators}
\end{figure}

The condensation renormalizes thus the propagators for the $f-f$, $c-c$ and $f-c$ channels. In fact, this equals summing the infinite series of the cumulant expansion due the condensation part of the interaction term (see Fig.-\ref{propagators}). 

\subsubsection{Derivation of the Luttinger-Ward functional}

Once we have properly handled the condensation part of the interaction, we can follow the strategy of Sec. \ref{SF_LW}.

The interaction term due to
hybridization fluctuations writes
 \bqa  \mathcal{S}_b = \frac{V}{\sqrt{N}} T^2\sum_{k, \sigma, \nu} \sum_{q \neq 0} \left (b_q f^\dagger_{\sigma \nu \, k+q} c_{\sigma \nu \, k} + H.c. \right ) , \eqa
 where we have extended the model to $N$ identical species of fermions by adding a channel index $\nu \in [1,N]$ to the fermionic operators and the $1/\sqrt{N}$ factor ensures a well-defined large-N limit.

\begin{figure}[h]
\center
\includegraphics[width=4.2 in, angle = 0]{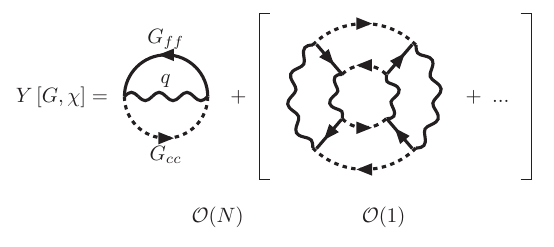}
\caption{ Leading skeleton diagrams participating to the LW functional $Y_b$ for the Kondo Breakdown model with dependence on $1/N$. The fermionic propagators correspond to $G_{ff}$ (straight line) and $G_{cc}$ (dashed line)  Green's functions of the zero order theory including the condensation part of the interaction. Both fermionic and bosonic propagators are fully dressed. The first diagram contains one fermionic loop carrying spin and channel quantum numbers and one pair of vertices, each of order $\mathcal{O}(1/\sqrt{N})$, so that it is of order $\mathcal{O}(N^2/N)=\mathcal{O}(N)$. The second diagram contains four loops and four pairs of vertices, so it is of order $\mathcal{O}(N^4/N^4)=\mathcal{O}(1)$.}
\label{skeleton_b}
\end{figure}

 Considering only the leading $\mathcal{O}(N)$ contribution to the LW functional shown in Fig.-\ref{skeleton_b}, and according to the general formula Eq.(\ref{LW}), we get the following expression for the free energy
\bqa \label{free_energy_KB} F_{LW}^{eff} =
F_F +  F_b + Y_{b}+  \Bigl( S N - \mathcal{B}^{2} \Bigr) \mu_{b} + u_b \frac{\mathcal{B}^4}{2}, \eqa
with
\bqa
 \label{LW_KB_b}
&F_F = -  \mbox{T Tr}\left [ \ln\left ( - G_0^{-1} + \Sigma \right ) + \Sigma G \right ]  \nn
&F_b =  \mbox{T Tr} \left  [ \ln\left ( - d_b^{-1} + \Pi_b \right ) + \Pi_b D_b \right ]  \\
& Y_b =  - 2 N V^2  \sum_{k} \sum_{q \neq 0} D_b(q) G_{cc}(k)G_{ff}(k+q) \nonumber \eqa

In (\ref{LW_KB_b}), $\Sigma=\left ( \begin{array}{cc}
\Sigma_{cc}& \Sigma_{cf}\\
\Sigma_{fc}& \Sigma_{ff}
\end{array} \right )$ and $G=\left ( \begin{array}{cc}
G_{cc}& G_{cf}\\
G_{fc}& G_{ff}
\end{array} \right )$ are the self-energy and full Green's matrices, respectively, of the fermions. They are related by the Dyson's equation
\bqa G^{-1}=G_0^{-1}-\Sigma. \label{dyson_f}\eqa
The same equation holds for holons
\bqa D_b^{-1}=d_b^{-1} - \Pi_b.\label{dyson_b}\eqa

\begin{figure}[ht]
 \center
\includegraphics[width=3.9 in, angle = 0]{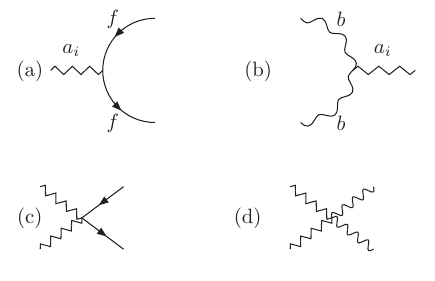}
\caption{Vertices due to the interaction of the gauge fields with holons and spinons.  Gauge propagator is represented by a zigzag line.}
\label{vertices_a}
\end{figure}

One can manipulate gauge fluctuations in the same way as the
above, where the gauge-coupling action, whose vertices are shown in Fig-\ref{vertices_a}, is given by \bqa &&
{\cal S}_{a}^{f} =\frac{1}{m_f} \sum_{k,q} \left |\k-\frac{\q}{2}\right | \left ( a_{q}f^\dagger_{\sigma k}f_{\sigma k-q}+H.c. \right )+ \frac{1}{2m_f}\sum_{k,q',q}  a^\dagger_{q'}a_{q'+q}f^\dagger_{\sigma k+q}f_{\sigma k}, \nn &&
 {\cal S}_{a}^{b} =\frac{1}{m_b} \sum_{q',q} \left |\q'-\frac{\q}{2}\right | \left ( a_{q}b^\dagger_{\sigma q'}b_{\sigma q'-q}+H.c. \right ) + \frac{1}{2m_b}\sum_{k,q',q}  a^\dagger_{q'}a_{q'+q}b^\dagger_{\sigma k+q}b_{\sigma k},  \eqa

\subsubsection{Introduction of gauge fluctuations}

Following exactly the same procedure for hybridization
fluctuations, one finds the following additional terms in Eq.(\ref{free_energy_KB}) \bqa
F_a&=&\mbox{T Tr} \left  [ \ln\left ( - d_a^{-1} + \Pi_a \right ) + \Pi_a D_a \right ] \nn
Y_a &= & -\frac{NT^2}{2}\sum_{k, q\neq 0} F(q,k) G_{ff}(k)D_a(q)G_{ff}(k+q)\nn
& & -\frac{T^2}{2}\sum_{q,q'}B(q,q')D_b(q)D_a(q')D_b(q+q'),
\eqa
where
\bqa
d_a^{-1}(\q, \Omega)&=&\frac{\Omega^2+ \q^2}{2 g^2} + \frac{\mathcal{B}^2}{2m_b}, ~~~
D_a^{-1} \equiv d_a^{-1} - \Pi_a .\nonumber
\eqa
\begin{figure}[ht]
\center
\includegraphics[width=4.1 in, angle = 0]{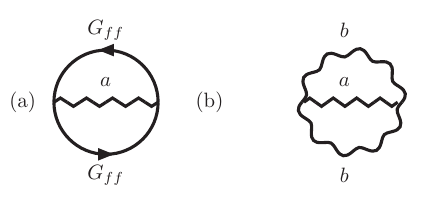}
\caption{ First order skeleton diagrams corresponding to $Y_a$. Fermionic and bosonic propagators are fully dressed. }
\label{skeleton_a}
\end{figure}

$F(k,q)$ and $B(q,q')$ are the current-gauge bare vertices for spinons and holons given respectively by
\bqa && F(k,q) \equiv \frac{1}{2}
\sum_{i,j=1}^{2} v_{i}^{f}\Bigl(\delta_{ij} -
\frac{q_{i}q_{j}}{q^{2}}\Bigr) v_{j}^{f} , ~~~ v_{i}^{f} =
\frac{k_{i} + q_{i}/2}{m_{f}} , \nn && B(k,q) \equiv \frac{1}{2}
\sum_{i,j=1}^{2} v_{i}^{b}\Bigl(\delta_{ij} -
\frac{q_{i}q_{j}}{q^{2}}\Bigr) v_{j}^{b} , ~~~ v_{i}^{b} =
\frac{k_{i} + q_{i}/2}{m_{b}} \nonumber ,\eqa

$Y_a$ corresponds to the contribution of the leading skeleton diagrams, due to interactions with the gauge field,  constructed with the fully dressed propagators of the spinons, the holons and the gauge fields (see Fig-\ref{skeleton_a}). 

\subsection{Eliashberg equations}

As for the HMM model, we can show that Eliashberg equations can be derived from the LW functional approach. Indeed, restricting ourselves to the leading $\mathcal{O}(N)$ terms of $Y=Y_b+Y_a$ shown in Fig-\ref{skeleton_b} and Fig-\ref{skeleton_a}, one can derive, in the same manner as  in the HMM case, the following expressions for the self-energies

\bqa
\Sigma_{cc}(k)&=& 2 V^2 T \sum_{q} D_b(q)G_{ff}(k+q) \nn
\Sigma_{ff}(k) &\equiv& \Sigma_{ff}^a + \Sigma_{ff}^b\nn
&=&T\sum_{q}F(k,q)G_{ff}(k+q)D_a(q) + 2 V^2 T \sum_q G_{cc}(k-q)D_b(q)\nn
\Pi_b(q) &\equiv& \Pi_b^a +  \Pi_b^{fc}\nn
&=& T \sum_{q'} B(q,q')D_a(q')D_b(q+q')\ +N V^2 T \sum_k G_{ff}(k+q)G_{cc}(k)\nn
\Pi_a(q)& \equiv& \Pi_a^f + \Pi_a^b\nn
&=& \frac{NT}{2}\sum_k F(k,q) G_{ff}(k)G_{ff}(k+q)+ \frac{T}{2}\sum_{q'}B(q',q)D_b(q)D_b(q+q')\nn
\label{self_KB_bis}
\eqa

\begin{figure}[ht]
\center
\includegraphics[width=4.2 in, angle = 0]{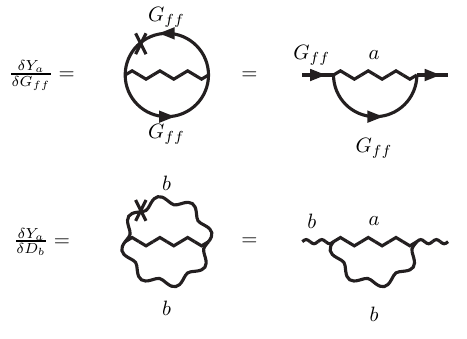}
\caption{Illustrating functional derivative of the gauge part $Y_a$ of LW functional with respect to Green's functions. The cross indicates the line that is cut by functional differentiation. Additional parts for the f-f channel and the holon self-energies are generated due to interactions with the gauge field.}
\label{functional_kb}
\end{figure}

We see that the gauge field induces an additional part in the self-energies of the spinons and the holons, which we notice $\Sigma_{ff}^a$ and $\Pi_b^a$ respectively (See Fig-\ref{functional_kb}). 

Equations (\ref{self_KB_bis}) are nothing but the Eliashberg equations for the KB model studied in \cite{Pepin_KBQCP}. It is found that this approximation becomes exact in the limit $N \rightarrow \infty$, provided we take into account the Fermi surface curvature\cite{Chubukov_LW_FM}. 

\subsection{Simplification of the Luttinger-Ward expression}

We can notice that 
\bqa
Y_b& \equiv&  - 2 N V^2  \sum_{k} \sum_{q \neq 0} D_b(q) G_{cc}(k)G_{ff}(k+q) \nn
& = & N T \sum_{k} \Sigma_{cc}(k)G_{cc} + \Sigma^b_{ff}(k)G_{ff} \label{Yb_FSE}\\
&=& - T \sum_{q} \Pi_{b}^{fc}(q) D_b(q)\label{Yb_BSE},
\eqa
and that
\bqa
Y_a &\equiv& -\frac{NT^2}{2}\sum_{k, q\neq 0} F(q,k) G_{ff}(k)D_a(q)G_{ff}(k+q)\nn
& & -\frac{T^2}{2}\sum_{q,q'}B(q,q')D_b(q)D_a(q')D_b(q+q')\nn
&=& NT \sum_k \Sigma^a_{ff}(k)G_{ff}(k) - \sum_{q} \Pi_b^a(q) D_b(q)\label{Ya_BbSE}\\
&=& NT \sum_k \Sigma^a_{ff}(k)G_{ff}(k) - \sum_{q} \Pi^b_a(q) D_a(q)\label{Ya_BaSE}
\eqa

Hence, we can simplify the expression of free energy for the KB model by introducing either equations (\ref{Yb_FSE}-\ref{Ya_BaSE}) into the expression of the free energy. As an example, if we use Eq. (\ref{Yb_BSE}) and (\ref{Ya_BaSE}) we get the following expression for free energy
\bqa \label{free_energy_KB_bis} F_{LW}^{eff} =
F_F +  F_b + F_{a}+  \Bigl( S N - \mathcal{B}^{2} \Bigr)( \mu_{b}+i a_0) + u_b \frac{\mathcal{B}^4}{2}, \nonumber \eqa
with\bqa
&F_F = -  \mbox{T Tr}\left [ \ln\left ( - G^{-1} \right ) + \Sigma^b G \right ]  \nn
&F_b =  \mbox{T Tr} \left  [ \ln\left ( - D_b^{-1} \right ) + \Pi_b^a D_b \right ]  \nn
&F_a =  \mbox{T Tr} \left  [ \ln\left ( - D_a^{-1} \right )  \right ]  
\label{LW_KB_bis} \eqa

In this expression, the fermionic contribution reduces to a Fermi liquid form (see  \ref{app:fermi_KB})
 $$F^F_{LW} \approx - \frac{\pi N\rho_{+}}{6}
T^{2} - \frac{\pi N\rho_{-}}{6} T^{2}, $$
where $\rho_{\pm}$ is the density of states of the upper(lower) hybridized band. The singular part of the free energy is given solely by the bosonic sector :  $F_s=F_{a}+F_b$.

We can make a further simplification considering that $\Pi^{fc}_b \gg \Pi_b^a$ \footnote{In the non-condensed phase, $\mathcal B = 0$, the contribution of the gauge fields to the boson propagator has been shown \cite{Pepin_KBQCP} to be dominant only for $T\leq E^*$  as it is then the only source of damping for the bosons. In the condensed phase with $\mathcal B\neq 0$, $\Pi_{fc}$ gets an additional term proportional to $\mathcal B^2$. (See Eq. (31)) and is thus leading.}, in which case the holon part is given by 
$$F_b = T \sum_q \ln \left ( -d_b^{-1}(q) + \Pi^{fc}(q) \right ).$$

\subsection{Thermodynamics}
Performing the energy and momentum integrals in the Eliashberg
equations [Eqs. (\ref{self_KB_bis})], we find Landau damping expressions for the holon and the gauge polarizations
 \bqa &&
\Pi_{b}^{fc}(q,i\Omega) = \frac{\gamma_{b}}{2m_{b}}
\frac{|\Omega|}{q} , \nn && \Pi_{a}^{f}(q,i\Omega) +
\Pi_{a}^{b}(q,i\Omega) = \frac{\gamma_{a}}{2m_{a}}
\frac{|\Omega|}{q}  ,  \eqa where 
$$\gamma_{b} =
\frac{2\pi}{v_{F}^{f}} \,\,\,\,\,\,\, \frac{\gamma_{a}}{m_a} =  \frac{N \pi}{m_f v_F^f} + \frac{\pi f_d}{m_b} $$ 
with $m_{b} = \frac{2}{N V^{2} \rho_{c}}$
and $f_d=\int \frac{d^{(d-1)}}{(2\pi)^(d-1)}\frac{q^2}{d}$ with a UV cut-off \cite{Pepin_KBQCP}. 
\newline

The singular contribution for the free energy is \bqa &&
f_{s}(\mu_{b}, T) = f_{c}(\mu_{b},T)  + T \sum_{q}\ln \Bigl( q^{2} + \gamma_{b}
\frac{|\Omega|}{q} + \Delta_{b}(\mu_{b},T) \Bigr) \nn && + T
\sum_{q}  \ln \Bigl( q^{2} +
\gamma_{a} \frac{|\Omega|}{q} + \Delta_{a}(\mu_{b},T) \Bigr) , \label{singular_KB}
\eqa where the condensation part $f_{c}(\mu_{b},T)$, holon mass
$\Delta_{b}(\mu_{b},T)$, and gauge-boson mass
$\Delta_{a}(\mu_{b},T)$ are given by \bqa && f_{c}(\mu_{b},T) = -
\mu_{b} \mathcal{B}^{2} + \frac{u_{b}}{2} \mathcal{B}^{4} , \nn &&
\Delta_{b}(\mu_{b},T) = - 2m_{b}(\mu_{b}) [\mu_{b} - 2 u_{b}
\mathcal{B}^{2}(\mu_{b},T) ] , \nn && \Delta_{a}(\mu_{b},T) =
\frac{m_{a}}{m_{b}(\mu_{b})} \mathcal{B}^{2}(\mu_{b},T) ,   \eqa
respectively. Note that the holon band mass depends on the
effective chemical potential since it is given by the electron
density of states. The coefficient in the gauge-boson mass is
given by $\frac{m_{a}}{m_{b}(\mu_{b}>0)} \approx \mathcal{O}(1)
\Bigl( \frac{V}{D}\Bigr)^{2}$, approximately.
\newline

An important remark is that we can determine self-consistently the condensation value $\mathcal{B}$ by the condition $\frac{\partial F_{LW}^{eff}}{\partial \mathcal{B}}=0$ (See \ref{app:self_B}). Beside the part obtained at the mean-field level, there are contributions due to hybridization and gauge fluctuation corrections. The Eliashberg framework allows then to refine the value of the condensation $\mathcal{B}$.

An explicit analytic expression for the singular part of the free energy is obtained after integration on frequencies and momenta in Eq.(\ref{singular_KB}). The details of this evaluation for $d=3, z=3$ and its results are given in \ref{app:singular_KB}. 

\subsection{Scaling of the free energy near the Kondo breakdown
quantum critical point}

Once we have the analytic expression for the singular part of the free energy, after integration of frequencies and momenta in Eq.(\ref{singular_KB}),  we can deduce its scaling expressions. As shown previously, this  part of the
free energy results from collective boson excitations associated
with hybridization and gauge
fluctuations. For each of these bosonic excitations, we associate a length scale for such
boson excitations, and the scaling form of the free energy near the Kondo Breakdown QCP reads \bqa &&
\hspace{-1cm}f_{s}(\xi_{b}^{-2},\lambda_{a}^{-2},T) = b_{b}^{-(d+z_{b})}
f_{b}(\xi_{b}^{-2} b_{b}^{1/\nu_{b}}, T b_{b}^{z_{b}})  +
b_{a}^{-(d+z_{a})} f_{a}(\lambda_{a}^{-2} b_{a}^{1/\nu_{a}}, T
b_{a}^{z_{a}}) . \eqa 

$f_{b(a)}(x,y)$ is an analytic regular
function for hybridization (gauge) fluctuations and $d$ is the space dimension. $\xi_{b} =
\Delta_{b}^{-1/2}$ is the correlation length for holons, and
$\lambda_{a} = \Delta_{a}^{-1/2}$ is the one for gauge bosons. In
particular, $\lambda_{a}$ may be considered as the penetration
depth in the superconductor. $b_{b}$ and $b_{a}$ are scaling
parameters for hybridization and gauge fluctuations, respectively \footnote{These scaling factors are in fact the same but may be associated with different dynamical exponents for different parts of the bosonic sector. Thus, we introduce subscripts for holons and gauge bosons to keep in mind this fact.}.
$\nu_{b(a)}$ is the correlation-length exponent of holons (gauge
bosons), and $z_{b(a)}$ is the dynamical exponent of holons (gauge
bosons). Here we have $z_{b} = z_{a}$ as shown in Eq. (47). Furthermore, $\nu_{b} = \nu_{a}=1/2$ as there are no quantum corrections to the usual Ornstein-Zernicke form of a static boson propagator within the Eliashberg treatment. These values for $\nu_{a,b}$ coincide with the mean-field value of the correlation length critical exponent in Hertz theory above its upper critical dimension.

Although two kinds of length scales are introduced, both scales
diverge at the same parameter point, $V =
V_{c}$ because they are related with each other via
Anderson-Higgs mechanism. In addition, we note that this
expression is applicable near the Kondo breakdown QCP, approaching
from the heavy-fermion side because we have considered properly and in a self-consistent way the effect of a finite condensation. 

Inserting $b_{b} = \xi_{b}^{2\nu}$ and $b_{a} =
\lambda_{a}^{2\nu}$ into the above scaling expression, we find
\bqa && f_{s}(\xi_{b}^{-2}, \lambda_{a}^{-2}, T)  =
\xi_{b}^{-2\nu(d+z)} f_{b}(1, T \xi_{b}^{2 \nu z}) +
\lambda_{a}^{-2\nu(d+z)} f_{a}(1, T \lambda_{a}^{2 \nu z}) . \label{scaling_KB_1}\eqa
This is our main result, derived from the microscopic
model based on the LW functional approach in the Eliashberg
framework. Now, one can understand thermodynamics near the Kondo
breakdown QCP based on this scaling free energy.

\section{Discussion and summary}
\label{discussion}
In this study we derived the scaling of free energy from a microscopic model  for two models of quantum criticality :  the standard
theoretical framework called the Hertz-Moriya-Millis (HMM) theory
and the strong coupling approach corresponding to the gauge theory. Fluctuation corrections
are taken into account systematically in the
Luttinger-Ward functional approach.  The Eliashberg framework allows to use the proper level of approximation to get, self-consistently, the correct scaling for
thermodynamics near the quantum critical point (QCP). We have shown that the singular part of the free energy for both models is due to the collective bosonic excitations, whereas the fermionic excitations give a Fermi Liquid contribution.

For the HMM theory, there exists one length scale associated with
the corresponding symmetry breaking, here spin-density-wave (SDW)
instability. This fact allows us to construct the scaling free
energy as a function of the spin-spin correlation length and
temperature for the SDW quantum transition. We derived the correct scaling expression using the Luttinger-Ward functional
approach in the Eliashberg framework.

For the gauge theory, there are additional collective excitations. These have 
nothing to do with the phase transition directly although they are affected by it. Such
collective modes turn out to be gauge fluctuations corresponding
to collective spin fluctuations in our context. An additional length scale,  associated with gauge
fluctuations, can appear. Indeed, considering that the Kondo breakdown transition is driven by condensation of holons, corresponding to the 
formation of an effective hybridization, the structure of the theory gives
rise to massive gauge fluctuations via the Anderson-Higgs
mechanism. This is the physical reason why the second length scale
appears in the gauge theory.

Because the two kinds of length scales, correlation length of
hybridization fluctuations and penetration depth of gauge
fluctuations, are deeply related via the Anderson-Higgs mechanism, they diverge at the Kondo breakdown QCP simultaneously.
However, the presence of the additional length scale leads to a different scaling expression for the thermodynamic
potential, compared with the HMM theory. We derived such a scaling expression using the Luttinger-Ward functional approach in the Eliashberg
framework. In the Eliashberg approximation, we showed that the
scaling expression of the free energy has two contributions
corresponding to each length scale, where each part contains only
one length scale.

In this paper we ignored vertex corrections, sometimes justified
but not always \cite{Chubukov_LW_FM}. Our path integral derivation
of the Luttinger-Ward functional gives a chance to extend the
Eliashberg framework, allowing vertex corrections. This can be achieved by going to higher orders of the cumulant expansion. In particular, if we
do the same job up to the fourth order, we expect that vertex
corrections will appear, satisfying the Bethe-Salpeter equation
for vertices \cite{Parquet}. It is an important future direction
to see how introduction of vertex corrections changes the scaling
expression of the Eliashberg approximation.
\newline

 This work is supported by the French National Grant ANR26ECCEZZZ.

\newpage

\appendix

\section{Derivation of the Luttinger-Ward functional up to second order in the interaction}
\label{app:derivation_LW}
The LW functional can be derived thoroughly using a cumulant expansion to the second order in the interaction term. Indeed, this term induces the following corrections to the bare action \bqa && \delta \mathcal{S}_{0} \approx - g^{2} T^4 \sum_{k, k'}\sum_{q, q'}  \Bigl[ \psi_{\alpha k}^{\dagger} \Bigl\langle
\psi_{\beta k} \psi_{\alpha' k'+q'}^{\dagger}
{\tau}^{n}_{\alpha\beta} S^{n}_{-q} {\tau}^{m}_{\alpha'\beta'}
S^{m}_{-q'} \Bigr\rangle_{c} \psi_{\beta' k'} \nn && + \Bigl\langle
\psi_{\beta k} \psi_{\alpha' k'+q'}^{\dagger}
{\tau}^{n}_{\alpha\beta} S^{n}_{-q} {\tau}^{m}_{\alpha'\beta'}
S^{m}_{-q'} \Bigr\rangle_{c} \Bigl\langle \psi_{\alpha k+q}^{\dagger}
\psi_{\beta' k'} \Bigr\rangle_{c}+  \psi_{\alpha k}^{\dagger} \Bigl\langle
\psi_{\alpha' k'+q'} \psi_{\beta' k'}^{\dagger}
{\tau}^{n}_{\alpha\beta} S^{n}_{-q} {\tau}^{m}_{\alpha'\beta'}
S^{m}_{-q'} \Bigr\rangle_{c} \psi_{\beta k} \nn && + \Bigl\langle
\psi_{\alpha' k'+q'} \psi_{\beta' k'}^{\dagger}
{\tau}^{n}_{\alpha\beta} S^{n}_{-q} {\tau}^{m}_{\alpha'\beta'}
S^{m}_{-q'}  \Bigr\rangle_{c} \Bigl\langle \psi_{\alpha k+q}^{\dagger}
\psi_{\beta k} \Bigr\rangle_{c}
 \Bigr] \nn && -
\frac{g^{2}}{2}T^4 \sum_{k, k'}\sum_{q, q'}  \Bigl[ S^{n}_{-q}
\Bigl\langle \psi_{\alpha k+q}^{\dagger}
{\tau}^{n}_{\alpha\beta}\psi_{\beta k}
\psi_{\alpha' k'+q'}^{\dagger}
{\tau}^{m}_{\alpha'\beta'}\psi_{\beta' k'} \Bigr\rangle_{c}
S^{m}_{-q'} \nn && + \Bigl\langle \psi_{\alpha k+q}^{\dagger}
{\tau}^{n}_{\alpha\beta}\psi_{\beta k}
\psi_{\alpha' k'+q'}^{\dagger}
{\tau}^{m}_{\alpha'\beta'}\psi_{\beta' k'} \Bigr\rangle_{c}
\Bigl\langle S^{n}_{-q} S^{m}_{-q'} \Bigr\rangle_{c} \Bigr] \nn && -
\frac{g^{2}}{2} T^4 \sum_{k, k'}\sum_{q, q'}\Bigl [\Bigl\langle
\psi_{\alpha k+q}^{\dagger} \psi_{\beta' k'} \Bigr\rangle_{c}
{\tau}^{n}_{\alpha\beta} {\tau}^{m}_{\alpha'\beta'}
\Bigl\langle S^{n}_{-q} S^{m}_{-q'} \Bigr\rangle_{c} \Bigl\langle
\psi_{\beta k}\psi_{\alpha' k'+q'}^{\dagger } \Bigr\rangle_{c}\nn &&+
\Bigl\langle
\psi_{\alpha k+q}^{\dagger} \psi_{\beta k} \Bigr\rangle_{c}
{\tau}^{n}_{\alpha\beta} {\tau}^{m}_{\alpha'\beta'} \Bigl\langle S^{n}_{-q} S^{m}_{-q'} \Bigr\rangle_{c} \Bigl\langle
\psi_{\alpha' k'+q'}^{\dagger } \psi_{\beta' k'}\Bigr\rangle_{c}\Bigr],\nn
\label{expansion} 
\eqa

 The fermionic and bosonic propagators are introduced as \bqa  G(k) \delta_{kk'}\delta_{\sigma \sigma'}&\equiv& -
\Bigl\langle \psi_{\sigma k} \psi_{\sigma' k'}^{\dagger}
\Bigr\rangle_c , \,\,\,\,\,\,\,\,
 \chi(q)\delta_{qq'} \equiv
\Bigl\langle S^{n}_{q} S^{n}_{-q'} \Bigr\rangle_c  \nonumber \eqa 
while the corresponding self-energies are \bqa   \Sigma(k)\delta_{kk'}\delta_{qq'}\delta_{\beta \alpha'} &\equiv& - g^{2} \Bigl\langle \psi_{\beta k} \psi_{\alpha' k'+q'}^{\dagger}
{\tau}^{n}_{\alpha\beta} S^{n}_{-q} {\tau}^{m}_{\alpha'\beta'}
S^{m}_{-q'} \Bigr\rangle_{c}  \nn 
&& -g^2   \Bigl\langle \psi_{\alpha' k'+q'} \psi_{\beta' k'}^{\dagger}
{\tau}^{n}_{\alpha\beta} S^{n}_{-q} {\tau}^{m}_{\alpha'\beta'}
S^{m}_{-q'} \Bigr\rangle_{c} ,\nn
\Pi(q)\delta_{qq'}
&\equiv& - g^{2} \Bigl\langle \psi_{\alpha k+q}^{\dagger}
{\tau}^{n}_{\alpha\beta}\psi_{\beta k}
\psi_{\alpha' k'+q'}^{\dagger}
{\tau}^{m}_{\alpha'\beta'}\psi_{\beta' k'} \Bigr\rangle_{c} . \nonumber
\eqa

\begin{figure}[ht]
\center
\includegraphics[width=3.4 in, angle = 0]{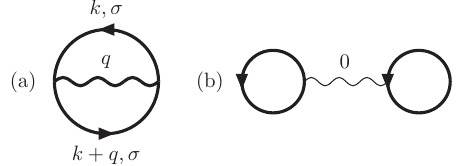}
\caption{The one loop closed diagrams for free energy in the SF model. The fermionic and bosonic propagators are bare}
\label{expansion_sf}
\end{figure}

The two last sums in (\ref{expansion}) corresponds to the two diagrams shown in Fig-\ref{expansion_sf} where the fermionic and bosonic propagators are bare. 

The prescription to get the LW functional at this order is to dress these propagators\cite{LW_Original} in the corresponding skeleton diagram, resulting in the following expression for the LW functional
 \bqa 
 \hspace{-1cm}Y = & N \frac{3g^{2}}{2}
T^2 \sum_{k, q}
G(k) \chi(q)G(k+q) + N^2 \frac{3g^2}{2}T^2 \chi(0) \sum_{k}G(k)\sum_{k'}G(k'). \eqa
In Fig-\ref{expansion_sf}, the diagram (b) will generate a static and uniform part in the self-energy and can thus be considered as a renormalization of the electron chemical potential.

\section{Fermionic contribution to free energy}
\subsection{Spin-fermion model }
\label{app:fermi_HMM}
The momentum dependence of the electron self-energy is shown to be
regular \cite{Chubukov_LW_FM}, we then replace the momentum with the Fermi
momentum $k_{F}$ : $\Sigma(i\omega) \equiv \Sigma(\k_F, i\omega)$. Then, the electron contribution to the free
energy of the SF model is given by \bqa
 \label{energy_electron}  F_{el}&& \equiv - N T \sum_{i\omega, k} \Bigl[\ln\Bigl\{ -
G_{0}^{-1}(k,i\omega) + \Sigma(i\omega) \Bigr\}  +
\Sigma(i\omega)G(k,i\omega) \Bigr]\nn
&& = - NT \sum_{i\omega, k}\Bigl[ \int_{0}^{1}du \,\,\partial_u \ln\Bigl\{- G_0^{-1}(k, i\omega) + u \Sigma(i\omega)\Bigr\}\nn
&&+ \frac{\Sigma(i\omega)}{G_0^{-1}(k, i\omega)-\Sigma(i\omega)}\Bigr] -NT\sum_{i\omega, k} \ln \Bigl\{-G_0^{-1}(k, i\omega)\Bigr\}\nn
\eqa
Here, the last term corresponds to the free fermion part giving raise to the Fermi liquid form of the free energy for electrons
\bqa
-NT \sum_{i\omega, k}\ln \Bigl\{ -G_0^{-1}(k, i\omega)\Bigr\} = -N\frac{\pi \rho_{F}}{6} T^2,
\eqa
where $\rho_{F}$ is the density of states
at the Fermi level. 
The first two terms of (\ref{energy_electron}) are shown to be vanishingly small in the low energy limit. Indeed, we have
\bqa
\delta F_{el} & \equiv & -NT \sum_{i\omega, k}\Bigl[ \int_{0}^{1}du \,\,\partial_u \ln\Bigl\{- G_0^{-1}(k, i\omega) + u \Sigma(i\omega)\Bigr\}\ +\frac{\Sigma(i\omega)}{G_0^{-1}(k, i\omega)-\Sigma(i\omega)}\Bigr]\nn
&=&-NT \sum_{i\omega, k} \int_0^1 du \Bigl[ -\frac{\Sigma(i\omega)}{G_0^{-1}(k,i\omega)-u\Sigma(i\omega)} + \frac{\Sigma(i\omega)}{G_0^{-1}(k,i\omega)-\Sigma(i\omega)} \Bigr]\nn
&=& -NT \sum_{i\omega, k}\frac{\Sigma(i\omega)^2}{G_0^{-1}(k,i\omega)-\Sigma(i\omega)}\int_0^1 du \frac{(1-u)}{G_0^{-1}-u\Sigma}.\nn
\eqa
Now, we can switch from the integration over momentum to that over energy as follows
$$ \sum_k ... \rightarrow \rho_{F}
\int_{-\Lambda}^{\Lambda} {d \epsilon } ... $$
where $\Lambda$ is an energy cut-off.

Integrating over $\epsilon$, we find that
\bqa
\hspace{-1cm}\delta F_{el} &&= - NT \rho_F \int_0^1 du \sum_{i \omega} \Sigma(i \omega) \Bigl[ \ln \Bigl(\frac{i \omega - \Sigma(i \omega)-\Lambda}{i \omega - \Sigma(i\omega) + \Lambda}\Bigr)- \ln \Bigl(\frac{i \omega - u \Sigma(i \omega)-\Lambda}{i \omega - u \Sigma(i\omega) + \Lambda}\Bigr) \Bigr]\nonumber
\eqa
For $|u \Sigma(i \omega)| \ll \Lambda$, the two last terms cancels and $\delta F_{el}$ vanishes. The electronic part of the free energy (\ref{free_energy_F_SF}) of the SF model reduces then to the Fermi liquid contribution
\bqa
\label{energy_electron_FL}
F_{el} = -N \frac{\pi \rho_F}{6}T^2.
\eqa

\subsection{Kondo breakdown theory}
\label{app:fermi_KB}

The fermionic sector in the KB model factorizes into an upper (+) and a lower (-) band whose dispersions are given by
$$E_{\mathbf{k}\pm} = \frac{1}{2}\left [ \epsilon_{\mathbf{k}} + \epsilon^{0}_{\mathbf{k}} \pm \sqrt{(\epsilon_{\mathbf{k}}-\epsilon^{0}_{\mathbf{k}})^2 + 4 V^2 \mathcal{B}^2} \right ] \,$$

The free energy for each band has a similar expression to Eq. (\ref{energy_electron_FL}) and reduces to a Fermi liquid form
$$F_{\pm} = - \frac{\pi
N\rho_{\pm} }{6} T^{2} ,$$ where $\rho_{\pm}$  is the density of states of the upper (lower) band at the Fermi level given by

$$\rho_{\pm}= \rho_{0}\left (\frac{\partial E_{\mathbf{k}\pm}}{\partial \epsilon_{\mathbf{k}}}\right )^{-1}_{|E_{\pm}=0}$$






\section{Momentum and frequency integral for the free energy of the Spin-fermion model}
\label{app:singular_SF}

Introducing $f(x) = \tan^{-1} x$, we have the following limits 
\bqa\hspace{-0.5cm} && f(x \ll 1)
\approx x , ~~~~~ f(x \gg 1) \approx \frac{\pi}{2} . \eqa
Then, the free energy expression of Eq. (\ref{singular_LW_SF}) can be cast according to
\bqa \hspace{-0.5cm}  f_{s}(\xi,T) &\approx- \frac{1}{2\pi^{3}}
\int_{0}^{\infty} d \nu \coth\Bigl(\frac{\nu}{2T}\Bigr) 
&\left[ \frac{\nu}{\Omega_{s}} \int_{q_r}^{\infty} d \tilde{q} \frac{\tilde{q}^{2} }{\xi^{-2}  +\tilde{q}^{2}} +
 \frac{\pi}{2}
\int_{0}^{qr} d \tilde{q}
\tilde{q}^{2} \right] ,
\eqa
where $q_r= \sqrt{ \frac{\nu}{\Omega_s}-\xi^{-2}}$. Then

\bqa   \hspace{-0.5cm}f_{s}(\xi,T) &=& - \frac{1}{2\pi^{3}}
\int_{0}^{\infty} d \nu \coth\Bigl(\frac{\nu}{2T}\Bigr) \Bigl[
\frac{\nu}{\Omega_{s}} \Bigl\{ \Lambda_{q} -\sqrt{\frac{\nu}{\Omega_{s}} - \xi^{-2}}  \nn & -& \xi^{-1} \Bigl(
\frac{\pi}{2} - \tan^{-1}\sqrt{\frac{\xi^{2}\nu}{\Omega_{s}} - 1}
\Bigr) \Bigr\} + \frac{\pi}{6} \Bigl(\frac{\nu}{\Omega_{s}} -
\xi^{-2}\Bigr)^{3/2} \Bigr]
\nn  &\approx& - \frac{1}{2\pi^{3}}
\frac{(2T)^{2}}{\Omega_{s}}\Lambda_{q} + \xi^{-5} f_{r}(T\xi^{2}),
\eqa where $\Lambda_{q}$ is a momentum cutoff and 
\bqa  f_{r}(T\xi^{2}) &=& \frac{1}{2\pi^{3}} \Bigl[ \frac{[2T\xi^{2}]^{2}}{\Omega_{s}}
\Bigl\{ \sqrt{\frac{2T\xi^{2}}{\Omega_{s}} - 1}  + \Bigl(
\frac{\pi}{2} - \tan^{-1}\sqrt{\frac{2T\xi^{2}}{\Omega_{s}} - 1}
\Bigr) \Bigr\} \nn &-& \frac{\pi}{6} [2T \xi^{2}] +
\Bigl(\frac{[2T\xi^{2}]}{\Omega_{s}} - 1 \Bigr)^{\frac{3}{2}}
\Bigr]. \nonumber
\eqa 
We see that the singular part of the free energy follows the
scaling relation shown in Eqs. (\ref{scaling_SF_1}) and (\ref{scaling_SF_2}).

\section{Momentum and frequency integral for the free energy in the Kondo breakdown scenario}
\label{app:singular_KB}
Let's consider the spectral representation of
Eq. (\ref{singular_KB})  \bqa && f_{s}(\mu_{r},T) =
f_{c}(\mu_{r},T) - \frac{1}{2\pi^{3}} \int_{0}^{\infty} {d \nu}
\coth\Bigl(\frac{\nu}{2T}\Bigr) \int_{0}^{\infty} d q q^{2} \nn &&
\Bigl\{ \tan^{-1}\Bigl( \gamma_{b} \frac{\nu}{q[q^{2} +
\Delta_{b}]} \Bigr) + \tan^{-1}\Bigl( \gamma_{a}
\frac{\nu}{q[q^{2} + \Delta_{a}]} \Bigr) \Bigr\} . \eqa 
Considering the approximation for $\tan^{-1} x$, the holon part  is cast, as previously, into two parts in the momentum integral  
\bqa 
f_{b}(\mu_{b},T) &=& - \frac{1}{2\pi^{3}} \int_{0}^{\infty} {d \nu}
\coth\Bigl(\frac{\nu}{2T}\Bigr) \int_{0}^{\infty} d q q^{2}
\tan^{-1}\Bigl( \gamma_{b} \frac{\nu}{q[q^{2} + \Delta_{b}]}
\Bigr) \nn & \approx& - \frac{1}{2\pi^{3}} \int_{0}^{\infty} {d
\nu} \coth\Bigl(\frac{\nu}{2T}\Bigr) \int_{q_{r}}^{\infty} d q
q^{2} \frac{\gamma_{b}\nu}{q[q^{2} + \Delta_{b}]} \nn & -&
\frac{1}{2\pi^{3}} \int_{0}^{\infty} {d \nu}
\coth\Bigl(\frac{\nu}{2T}\Bigr) \int_{0}^{q_{r}} d q q^{2}
\frac{\pi}{2}, \nonumber \eqa 
where $q_{r}$ is a characteristic momentum determined by the equation \bqa &&
\frac{\gamma_{b} \nu}{q_{r}[q_{r}^{2} + \Delta_{b}]} = 1
\rightarrow q_{r}^{3} + \Delta_{b} q_{r} - \gamma_{b} \nu = 0.\nonumber
\eqa
 The solution of the latter is given by \bqa \hspace{-0.8cm}&& q_{r} = -
\frac{(2/3)^{1/3} \Delta_{b}}{\Bigl(9 \gamma_{b} \nu + \sqrt{12
\Delta_{b}^{3} + 81 (\gamma_{b} \nu)^{2}}\Bigr)^{1/3}}  +
\frac{\Bigl(9 \gamma_{b} \nu + \sqrt{12 \Delta_{b}^{3} + 81
(\gamma_{b} \nu)^{2}}\Bigr)^{1/3}}{2^{1/3}3^{2/3}} , \eqa
which is definitely positive.

Then 
\bqa \hspace{-1.5cm}f_b(\mu_b,T) &=& - \frac{1}{4\pi^{3}} \int_{0}^{\infty} {d
\nu} \coth\Bigl(\frac{\nu}{2T}\Bigr)
\int_{q_{r}^{2}}^{\Lambda_{q}^{2}} d x \frac{\gamma_{b}\nu }{x +
\Delta_{b} } - \frac{1}{12\pi^{2}} \int_{0}^{\infty} {d
\nu} \coth\Bigl(\frac{\nu}{2T}\Bigr) q_{r}^{3} \nn \hspace{-1.5cm} & \approx &-
\frac{1}{4\pi^{3}} \int_{0}^{\infty} {d \nu}
\coth\Bigl(\frac{\nu}{2T}\Bigr) \gamma_{b} \nu \ln \Bigl(
\frac{\Lambda_{q}^{2} q_{r}}{\gamma_{b} \nu} \Bigr) -
\frac{1}{12\pi^{2}} \int_{0}^{\infty} {d \nu}
\coth\Bigl(\frac{\nu}{2T}\Bigr) \Bigl( - \Delta_{b} q_{r} +
\gamma_{b} \nu \Bigr) , \nn  \eqa
where the momentum cutoff
$\Lambda_{q}$ is taken much larger than the holon mass, i.e.,
$\Lambda_{q}^{2} \gg \Delta_{b}$.

The frequency integral can be performed approximately, given by
\bqa  \hspace{-1.5cm}f_{b}(\mu_{r},T) &=& - \frac{1}{4\pi^{3}} \int_{0}^{\infty}
{d \nu} \coth\Bigl(\frac{\nu}{2T}\Bigr) \gamma_{b} \nu \ln \Bigl(
\frac{\Lambda_{q}^{2} q_{r}}{\gamma_{b} \nu} \Bigr)  -
\frac{1}{12\pi^{2}} \int_{0}^{\infty} {d \nu}
\coth\Bigl(\frac{\nu}{2T}\Bigr) \Bigl( - \Delta_{b} q_{r} +
\gamma_{b} \nu \Bigr) \nn & \approx & - \frac{1}{4\pi^{3}} \Bigl\{
\int_{0}^{2T} {d \nu} \frac{2T}{\nu} + \int_{2T}^{\Lambda_{\nu}} d
\nu  \Bigr\} \gamma_{b} \nu \ln \Bigl( \frac{\Lambda_{q}^{2}
q_{r}[\Delta_{b},\nu]}{\gamma_{b} \nu} \Bigr) \nn && -
\frac{1}{12\pi^{2}} \Bigl\{ \int_{0}^{2T} {d \nu} \frac{2T}{\nu} +
\int_{2T}^{\Lambda_{\nu}} d \nu  \Bigr\} \Bigl( - \Delta_{b} q_{r}
+ \gamma_{b} \nu \Bigr) \nn & \approx & - \frac{1}{4\pi^{3}}
\int_{0}^{2T} {d \nu} \frac{2T}{\nu} \gamma_{b} \nu \ln \Bigl(
\frac{\Lambda_{q}^{2} q_{r}[\Delta_{b},\nu]}{\gamma_{b} \nu}
\Bigr)  - \frac{1}{12\pi^{2}} \int_{0}^{2T} {d \nu}
\frac{2T}{\nu} \Bigl( - \Delta_{b} q_{r} + \gamma_{b} \nu \Bigr)
\nn & \approx &- \frac{1}{4\pi^{3} \gamma_{b}} (2\gamma_{b}T)^{2}
\ln \Bigl( \frac{\Lambda_{q}^{2} q_{r}[\Delta_{b},2T]}{2\gamma_{b}
T} \Bigr)  - \frac{1}{12\pi^{2}\gamma_{b}} (2\gamma_{b}T)
\Bigl( - \Delta_{b} q_{r}[\Delta_{b},2T] + 2 \gamma_{b} T \Bigr) ,\nonumber
\eqa
 where
 \bqa \hspace{-1.5cm} q_{r}[\Delta_{b},2T] &=& - \frac{(2/3)^{1/3}
\Delta_{b}}{\Bigl(9 [2\gamma_{b}T] + \sqrt{12 \Delta_{b}^{3} + 81
(2\gamma_{b} T)^{2}}\Bigr)^{1/3}}  + \frac{\Bigl(9
[2\gamma_{b} T] + \sqrt{12 \Delta_{b}^{3} + 81 (2\gamma_{b}
T)^{2}}\Bigr)^{1/3}}{2^{1/3}3^{2/3}}  . \nn \eqa

For the gauge-fluctuation part, exactly the same procedure is
performed, and the result holds provided that the subscript $b$ is replaced with $a$.


\section{Introduction of dangerously irrelevant variables}
\label{app:corrections_scaling}
In this appendix, we show how the presence of a dangerously irrelevant variable can affect the naive scaling relations obtained in (\ref{scaling_SF_1}) and (\ref{scaling_KB_1}) 

We recall that Millis \cite{HMM} showed in this case, using perturbative RG, that the control parameter $\delta = \xi_0^{-2}$ is renormalized according to
\be \delta_r = \delta + u \frac{C}{z+d-2},\label{control_parameter} \ee
where C is a constant, and the correlation length $\xi$ gets a temperature dependence

\be \xi^{-2} = \delta_r + g(d, z) u T^\frac{d+z-2}{z},\label{app:correction_scaling} \ee
where $g$ is a function depending on the dimension $d$ and the dynamical exponent $z$.

Accordingly, the naive scaling shown in Eq.(\ref{scaling_SF_1}) is invalidated and the effect of the dangerously irrelevant parameter $u$ must be incorporated into a generalized scaling form 
\be f(\xi^{-2}, T, u) = b^{-(d+z)}f(\xi^{-2} b^{1/\nu}, Tb^z, ub^{d+z-4}).\label{app:generalized_scaling}\ee

\begin{figure}[ht]
\centering
\includegraphics[width=3.2 in]{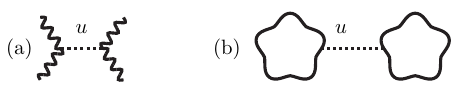}
\includegraphics[width=1.2 in]{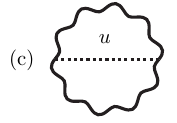}
\caption{(a) a $\phi^4$ vertex with a constant coefficient $u$, (b) and (c) are the first loop diagrams generated by the vertex (a) where the bosonic lines are fully dressed.}
\label{correction_scaling}
\end{figure}

Let's consider then a $\phi^4$ term with a constant coefficient $u$ in the SF model (\ref{SFM}), as in the Hertz-Millis theory, within our method. The corresponding vertex is shown in Fig.- \ref{correction_scaling}-(a) below.  At the first loop level, this quartic term generates the diagrams shown in Fig.- \ref{correction_scaling}-(b) and (c), where the bosonic propagators are full. These are of order $\mathcal{O}(1)$ and are thus sub-leading with respect to the diagram of order $\mathcal{O}(N)$ first considered in the LW functional shown in Fig.- \ref{Y_SF}.
\newline

 If these are included in the LW functional, the bosonic self-energy, generated by variation of the free energy with respect to the bosonic Green's function as explained in section \ref{ss:eliashberg_SF}, gets a constant contribution from the diagram (a). This is a renormalization of the bosonic chemical potential as in Eq.(\ref{control_parameter}).  The diagram (b) is easily shown to give corrections to scaling to the correlation length $\xi$ as
$$\xi^{-2}(T)-\xi^{-2}(0)\propto T^{\frac{d+z-2}{z}},$$
consistent with (\ref{app:correction_scaling})

We see that our method allows to account self-consistently for the effect of the parameter $u$. Rigorously, this should be considered in the expression of the free energy. Our purpose in the main text is only to show how an analytic expression for free energy can be obtained in a self-consistent way.
\newline

Similar considerations should hold for the Kondo breakdown scenario as well.  However, in practice, the naive scaling for the free energy can still be used to fit experimental data. This has been done in one of our previous papers \cite{Kim_Adel_Pepin} in the case of the Kondo Breakdown scenario to account for the anomalous exponent of the Gr\"uneisen ratio of the compound YbRh$_2$(S$_{0.95}$Ge$_{0.05}$)$_2$ The naive scaling for free energy gives a rather good agreement with the experiment in this case.

\section{Self-consistent equation for $\mathcal{B}$}
\label{app:self_B}
Minimizing the free energy Eq.\ref{free_energy_KB_bis} with respect to $\mathcal{B}$, we get the following expression
 \bqa \hspace{-2.5cm} 0 = H_{MF}(\mathcal{B})  + T \sum_{i\Omega } \int
\frac{d^{d} q}{(2\pi)^{d}} \frac{8 m_{b} u_{b} \mathcal{B}}{q^{2}
+ \gamma_{b} \frac{|\Omega|}{q} - 2m_{b} [\mu_{b} - 2 u_{b}
\mathcal{B}^{2}]} + T \sum_{i\Omega } \int \frac{d^{d}
q}{(2\pi)^{d}} \frac{\frac{2m_{a}}{m_{b}} \mathcal{B}}{q^{2} +
\gamma_{a} \frac{|\Omega|}{q} + \frac{m_{a}}{m_{b}}
\mathcal{B}^{2} } ,\nn \eqa
where $H_{MF}(\mathcal{B})=0$ determines the mean-field value of the condensation $\mathcal{B}$. Fluctuations of the holon and the gauge fields result in additional terms in the self-consistent equation for $\mathcal{B}$.

The holon part is evaluated as follows
\bqa \hspace{-1.5 cm} T \sum_{i\Omega }
\int \frac{d^{3} q}{(2\pi)^{3}} \frac{1}{q^{2} + \gamma_{b}
\frac{|\Omega|}{q} + \Delta_{b}} &= \int \frac{d^{3} q}{(2\pi)^{3}}
\int_{-\infty}^{\infty} d \nu \Bigl( - \frac{1}{\pi}\Bigr)
\frac{\gamma_{b} \nu /q}{(q^{2} + \Delta_{b})^{2} +
(\gamma_{b}\nu)^{2}/q^{2}}   T \sum_{i\Omega} \frac{1}{i\Omega -
\nu} \nn 
 &= \frac{1}{2\pi^{3}} \int_{0}^{\infty} d\nu
\coth\Bigl(\frac{\nu}{2T}\Bigr) \int_{0}^{\infty} d q
\frac{\gamma_{b} \nu q^{3}}{q^{2}(q^{2} + \Delta_{b})^{2} +
(\gamma_{b}\nu)^{2} } \nn
  &= \frac{1}{4\pi^{3}}
\int_{0}^{\infty} d\nu \coth\Bigl(\frac{\nu}{2T}\Bigr)
\int_{0}^{\infty} d x \frac{\gamma_{b} \nu x}{x(x +
\Delta_{b})^{2} + (\gamma_{b}\nu)^{2} } \nn 
& \approx
\frac{1}{4\pi^{3}} \int_{0}^{\infty} d\nu
\coth\Bigl(\frac{\nu}{2T}\Bigr)
\int_{Max[\Delta_{b},(\gamma_{b}\nu)^{2/3}]}^{\infty} d x
\frac{\gamma_{b} \nu x}{x^{3}} \nn  &= \frac{1}{4\pi^{3}}
\int_{0}^{\infty} d\nu \coth\Bigl(\frac{\nu}{2T}\Bigr)
\frac{\gamma_{b} \nu }{Max[\Delta_{b},(\gamma_{b}\nu)^{2/3}]} \nn
 &\approx \frac{1}{4\pi^{3}\gamma_{b}}  \frac{(2\gamma_{b}T)^{2}
}{Max[\Delta_{b},(2\gamma_{b}T)^{2/3}]}, \eqa
where the $Max$ function is defined as $Max[A,B] = A$ when $A
\geq B$.  

The gauge part is evaluated in the same way and the analytic expression for the self-consistent
equation of $\mathcal{B}$ writes \bqa && 0 = H_{MF}(\mathcal{B})  + \frac{m_{b} u_{b} }{\pi^{3}\gamma_{b}}
\frac{\mathcal{B}(2\gamma_{b}T)^{2} }{Max[-2m_{b} (\mu_{b} - 2 u_{b}
\mathcal{B}^{2}),(2\gamma_{b}T)^{2/3}]} \nn && +
\frac{m_{a}/m_{b}}{4\pi^{3}\gamma_{a}} \frac{\mathcal{B}(2\gamma_{a}T)^{2}
}{Max[\frac{m_{a}}{m_{b}} \mathcal{B}^{2},(2\gamma_{a}T)^{2/3}]} 
\eqa 
\newline

\end{document}